\documentclass[prd,
,superscriptaddress,
nofootinbib,%
tightenlines ]{revtex4}
\usepackage{epsfig}
\usepackage[colorlinks=true,linkcolor=blue,urlcolor=blue,citecolor=blue]{hyperref}
\usepackage{dsfont}
\usepackage{amsmath}
\usepackage{amsfonts}
\usepackage{amssymb}
\usepackage{bm}

\newcommand{\ben}{\begin{displaymath}}
\newcommand{\een}{\end{displaymath}}
\newcommand{\be}{\begin{equation}}
\newcommand{\ee}{\end{equation}}
\newcommand{\bea}{\begin{eqnarray}}
\newcommand{\eea}{\end{eqnarray}}

\newcommand{\nn}{\nonumber \\ }

\begin{document}
\title{On the definition of electromagnetic local spatial densities for composite \\spin-$1/2$ systems}
\author{J.~Yu.~Panteleeva}
  \affiliation{Institut f\"ur Theoretische Physik II, Ruhr-Universit\"at Bochum,  D-44780 Bochum,
 Germany}
\author{E.~Epelbaum}
 \affiliation{Institut f\"ur Theoretische Physik II, Ruhr-Universit\"at Bochum,  D-44780 Bochum,
 Germany}
\author{J.~Gegelia}
 \affiliation{Institut f\"ur Theoretische Physik II, Ruhr-Universit\"at Bochum,  D-44780 Bochum,
 Germany}
 \affiliation{High Energy Physics Institute, Tbilisi State
University, 0186 Tbilisi, Georgia}
\author{U.-G.~Mei\ss ner}
 \affiliation{Helmholtz Institut f\"ur Strahlen- und Kernphysik and Bethe
   Center for Theoretical Physics, Universit\"at Bonn, D-53115 Bonn, Germany}
 \affiliation{Institute for Advanced Simulation, Institut f\"ur Kernphysik
   and J\"ulich Center for Hadron Physics, Forschungszentrum J\"ulich, D-52425 J\"ulich,
Germany}
\affiliation{Tbilisi State  University,  0186 Tbilisi,
 Georgia}
 
\date{\today}
\begin{abstract}
An unambiguous definition of the electromagnetic spatial densities for  a spin-1/2 system
is proposed and worked out in the zero average momentum frame and in moving frames. The obtained
results are compared with the traditional definition of the densities in
terms of the three-dimensional Fourier transforms of the
electromagnetic form factors in the Breit frame.
   
\end{abstract}

\maketitle

\section{Introduction}

The three-dimensional  Fourier transform of the
charge form factor in the Breit frame is often interpreted as the electric charge density
of the corresponding hadron, owing to the seminal papers on
electron-proton scattering by Hofstadter, Sachs and others in the 60s
of the last century
\cite{Hofstadter:1958,Ernst:1960zza,Sachs:1962zzc}.  
Similar interpretations have also been proposed for the Fourier transforms of
the gravitational form factors and for various local distributions 
\cite{Polyakov:2002wz,Polyakov:2002yz,Polyakov:2018zvc}.   

Despite common belief, the identification of spatial density
distributions with the Fourier transform of the corresponding form
factors in the Breit frame suffers from conceptual issues as
repeatedly pointed out in the literature
\cite{Burkardt:2000za,Miller:2007uy,Miller:2009qu,Miller:2010nz,Jaffe:2020ebz,Miller:2018ybm,Freese:2021czn}.   
In Ref.~\cite{Jaffe:2020ebz}, it was shown on the example of a spin-$0$ system
that the above-mentioned traditional expression for the
charge density in terms of the Breit frame distribution follows only
in the static limit of an infinitely heavy particle. On the other
hand, doubt has been raised that local density distributions can even be
defined unambiguously, i.e.~independently of the form 
of the wave packet, for systems whose Compton wavelength is of the
order of or larger than the charge radius (defined via the derivative
of the form factor at zero momentum transfer) \cite{Jaffe:2020ebz}.

The issue of a proper definition of the spatial distributions of matrix elements of local
operators has attracted much attention in the last few years
\cite{Lorce:2020onh,Lorce:2022cle,Guo:2021aik,Lorce:2018egm,Freese:2021mzg}.
In the recent work of Ref.~\cite{Epelbaum:2022fjc},  the proper
definition of the charge density
has been revisited
for a spin-$0$ system.\footnote{We thank Cedric~Lorc\'e for pointing out that similar results
have been published long ago in Ref.~\cite{Fleming:1974af}. } Closely following the
logic of Ref.~\cite{Jaffe:2020ebz},
the charge density has been defined unambiguously in the zero average momentum frame (ZAMF)  of
the system by using spherically symmetric sharply localized wave
packets without invoking any approximations.\footnote{Under ZAMF we mean a Lorentz frame in
 which the expectation value of the 
three-momentum for the state, specified by the given packet, is zero. 
For wave packets with a sharp localization around an eigenstate of the four-momentum operator, the ZAMF coincides with the rest-frame of the system.} The definition has also been generalized to moving 
frames, and it was shown that  in the infinite-momentum frame (IMF), the charge density turns
into the well-known two-dimensional distribution in the transverse
plane times the delta-function in the longitudinal direction. 

In the current paper we work out the details of our new definition of the electromagnetic densities 
for spin-$1/2$ systems. Analogously to the spin-$0$ case, we consider
sharply localized wave packets and obtain local spatial distributions
for the ZAMF as well as for moving frames. 

Our work is organized as follows, in Section~\ref{rest} we consider the spatial distributions
of a spin-1/2 system in the ZAMF. Section~\ref{moving} is devoted to the moving frames.
In Section~\ref{sec:Disc}, we discuss the interpretation of the novel
density distributions and comment on the proton radius controversy.
We end with a summary in  Section~\ref{summary}.

\section{The electromagnetic densities of a spin-1/2 system in the ZAMF}
\label{rest}

We start with considering the electromagnetic densities of a spin-$1/2$ system in its ZAMF.  
We choose the four-momentum eigenstates
$|p,s\rangle$ characterizing our system to be normalized as
\begin{equation}
\langle p',s'|p,s\rangle = 2 E (2\pi )^3 \delta_{s's}\delta^{(3)} ({\bf p'}-{\bf p})\,,
\label{NormStateN}
\end{equation}
where $p=(E,{\bf p})$, with $E=\sqrt{m^2+{\bf p}^2}$ and $m$ the particle's mass. 
These states are also eigenstates of the charge operator given by $\hat Q=\int d^3{r} \, \hat
j^0({\bf r},0)$, where $\hat j^\mu({\bf r},0)$ is the electromagnetic current 
operator at $t=0$ in the  Heisenberg picture.

The matrix elements of  the electromagnetic current operator 
between momentum eigenstates of a spin-$1/2$ system can be parameterized in terms of two form factors:
\begin{eqnarray}
\langle p',s' | \hat j^\mu ({\bf r} ,0)| p ,s\rangle & = & e^{ - i({\bf p'}-{\bf p})\cdot {\bf r}} \, \bar u(p',s') \left[ \gamma^\mu F_1(q^2) 
+  \frac{1}{2}i \sigma^{\mu\nu} q_\nu \, F_2(q^2) \right] u(p,s) ,
\label{eqMEN}
\end{eqnarray} 
with $p$ and $s$ ($p'$ and $s'$) denoting the momentum and the polarization of the initial (final) states,
respectively. The Dirac spinors are normalized as $ \bar u(p,s') u(p,s) = 2 m \delta_{s's}$.  
The momentum transfer is given by $q = p' - p$, and $F_1(q^2)$ and $F_2(q^2)$ are the Dirac and the
Pauli form factors, respectively. These are normalized as $F_1(0)=1$ and $F_2(0)=\kappa/m$,
with $\kappa$ being the anomalous magnetic moment of the spin-$1/2$ particle. We point out  the
difference in the normalization of the Pauli form factor compared to the literature.

To define local electromagnetic densities we calculate the matrix element of the current
operator in a state localized in coordinate  space and take the size of the localization
much smaller than all length scales characterizing the system.  
A normalizable Heisenberg-picture state with the center-of-mass position ${\bf X}$ can
be written in terms of a wave packet as
\begin{equation}
|\Phi, {\bf X},s \rangle = \int \frac{d^3 {p}}{\sqrt{2 E (2\pi)^3}}  \, \phi(s,{\bf p})
\, e^{-i {\bf p}\cdot{\bf X}} |p ,s \rangle,  
\label{statedefN2}
\end{equation}
where from the normalization condition it follows that  
\begin{equation}
\int d^3 {p} \,  | \phi(s,{\bf p})|^2 =1\,.  
\label{normN}
\end{equation}
We {\it define} the density distributions in
the ZAMF of the system by employing spherically symmetric wave
packets with spin-independent profile functions  $\phi(s, {\bf p}) = \phi({\bf p}) = \phi(|{\bf p}|)$. 
For later convenience, we introduce a dimensionless profile function
$\tilde \phi $ via 
\begin{equation}
\phi({\bf p}) = R^{3/2} \, \tilde \phi(R  {\bf p})\,,
\label{packageFormN}
\end{equation} 
where $R$ characterizes the size of the wave packet such that sharp localization corresponds to
small values of $R$.
The current density distribution for the state defined in Eq.~(\ref{statedefN2}) takes the form
\begin{equation}
   \langle \Phi, {\bf X},s' | \hat
                                   j^\mu ({\bf r}, 0 ) | \Phi, {\bf X},s
                                   \rangle 
  = \int \frac{d^3 {p} \, d^3 {p}'}{(2\pi)^3 \sqrt{4 E
    E'}}
 \bar u(p',s') \left[ \gamma^\mu F_1(q^2)  
+   \frac{i \sigma^{\mu\nu} q_\nu}{2} F_2(q^2) \right] u(p,s)    
     \phi^\star({\bf p'}) 
                                 \phi({\bf p})  e^{i {\bf q}\cdot ({\bf X} - {\bf r})} .
\label{rhointN}
\end{equation}
Without loss of generality we can take ${\bf  X} = 0$. In terms of the momenta 
${\bf P}=({\bf p}' + {\bf p})/2$ and $ {\bf q}$ 
the current density distribution takes the form
\begin{eqnarray}
  j^\mu_\phi({\bf r}) &\equiv &
 \langle \Phi, {\bf 0},s' | \hat
                                   j^\mu ({\bf r}, 0 ) | \Phi, {\bf 0},s
                                \rangle  \nn
                                &=&
                                \int \frac{d^3 {P} \, d^3 {q}}{(2\pi)^3 \sqrt{4 E E'}}\,
\, \bar u(p',s') \left[ \gamma^\mu F_1((E-E')^2- {\bf q}^2) +   \frac{i \sigma^{\mu\nu} q_\nu}{2} \, F_2((E-E')^2- {\bf q}^2) \right] u(p,s) \nn
                &\times& \phi\bigg( {\bf P} -
\frac{\bf q}{2}\bigg) \, \phi^\star\bigg( {\bf P} +\frac{\bf q}{2}\bigg)  \, e^{ - i {\bf q}\cdot {\bf  r}} ,
\label{rhoint2N}
\end{eqnarray}
where $E=\sqrt{m^2+ {\bf P}^2 - {\bf P}\cdot {\bf q} +{\bf q}^2/4 } $
and $E'=\sqrt{m^2+ {\bf P}^2 + {\bf P}\cdot {\bf q} +{\bf q}^2/4 } $.

The standard definition of the current density in terms of the form factors in
the Breit frame, $F_i(q^2) = F_i (-{\bf q}^2)$, which we will refer
to as ``naive'' following the terminology of Ref.~\cite{Jaffe:2020ebz}, is obtained 
by first approximating  the integrand in Eq.~(\ref{rhoint2N}) by the
two leading terms in the $1/m$-expansion and
subsequently localizing the wave packet by taking the limit $R \to 0$
\cite{Miller:2018ybm,Jaffe:2020ebz}.
The $1/m$-expansion of the integrand leads to 
the following expressions:
\begin{eqnarray}
J^0_{\phi, \, \rm naive}({\bf r}) & \equiv&  \int \frac{d^3 {P} \, d^3 {q}}{(2\pi)^3} \, \left\{ F_1(- {\bf q}^2) 
-  F_2(- {\bf q}^2) \left[\frac{ {\bf q}^2}{4 m} - \frac{i}{2 m}\, {\bf q}\cdot ({\bf P}\times{\bm\sigma}) \right] \right\}  
                \phi\bigg({\bf P} -
\frac{\bf q}{2}\bigg) \, \phi^\star\bigg({\bf P} +\frac{\bf q}{2}\bigg)  \, e^{ - i {\bf q}\cdot {\bf  r}} , \nn 
{\bf J}_{\phi, \, \rm naive}({\bf r}) & \equiv & \frac{1}{m}\int \frac{d^3 {P} \, d^3 {q}}{(2\pi)^3}\,
\left[ {\bf P}\, F_1(- {\bf q}^2)  
-  \frac{F_1(- {\bf q}^2)+m F_2(- {\bf q}^2) }{2} \, i \, {\bf q}\times{\bm \sigma} \right]  
                \phi\bigg({\bf P} -
\frac{\bf q}{2}\bigg) \, \phi^\star\bigg( {\bf P} +\frac{\bf q}{2}\bigg)  \, e^{ - i {\bf q}\cdot {\bf  r}} \,.
\label{rhoint2NRN}
\end{eqnarray} 
Here and in what follows, we use $J^\mu$ instead of $j^\mu$ to indicate
that these densities are written as operators in spin space rather
than the corresponding matrix elements. 
The $R\to 0$ limit of the expressions in Eq.~(\ref{rhoint2NRN}) can be calculated without specifying the 
form factors and the profile function
$\phi ({\bf p} )$ using the methods developed to analyze loop
integrals in quantum field
theory \cite{Gegelia:1994zz,Beneke:1997zp}.
For $F_1\left( q^2\right)$ and $F_2\left( q^2\right)$ decreasing at  large $q^2$ faster than
$1/q^2$  and $1/(q^2)^2$, respectively, by using the method of dimensional counting
\cite{Gegelia:1994zz}, the only non-vanishing
contribution to $J^\mu_{\phi, \, \rm naive}({\bf r})$  in the $R\to 0$ limit is obtained by
substituting ${\bf P}= \tilde {\bf P}/R$, expanding the integrands in Eq.~(\ref{rhoint2NRN}) in $R$
around $R=0$ and keeping up to the zeroth order terms.  Doing so we obtain
\begin{eqnarray}
{ J}^0_{\rm naive}({\bf r})  \; \equiv \;  \lim_{R \to 0}  J^0_{\phi, \, \rm naive}({\bf r}) &=& 
\int \frac{d^3 {\tilde P} \, d^3 {q}}{(2\pi)^3} \,  \left\{ F_1(- {\bf q}^2) -    F_2(- {\bf q}^2) \left[\frac{ {\bf q}^2}{4 m} - \frac{1}{R}\,\frac{i}{2 m}\, {\bf q}\cdot ({\bf \tilde P} \times {\bm \sigma}) \right] \right\}  
                 |\tilde\phi({ {\bf \tilde  P}})|^2  \, e^{ -i {\bf q}\cdot {\bf  r}} \nn [3pt]
    &=&    \int \frac{d^3 {q}}{(2\pi)^3}\, e^{ - i {\bf
        q}\cdot {\bf  r}}  \, G_E (- {\bf q}^2 ) \nn [3pt]
& \equiv & \rho_{\rm naive}^{\rm ch} (r)
     \,    , \nn  [3pt]
{{\bf  J}}_{\rm naive}({\bf r})  \; \equiv \;  \lim_{R \to 0}   {\bf J}_{\phi, \, \rm naive}({\bf r}) &=&  \int \frac{d^3 {\tilde P} \, d^3 {q}}{(2\pi)^3}\,
\, \left\{ \frac{{\bf \tilde P}}{m R}\, F_1(- {\bf q}^2) -   \frac{ i  \, {\bf q} \times \bm{\sigma}  }{2 m} \, \left[ F_1(- {\bf q}^2) + m F_2(- {\bf q}^2)  \right] \right\}  
                 |\tilde\phi({
  {\bf \tilde  P}})|^2  \, e^{ - i {\bf q}\cdot {\bf  r}} \nn [3pt]
& =&  \frac{ {\bm
     \nabla_{\bf r}} \times  {\bm \sigma} }{2m} \int \frac{d^3 {q}}{(2\pi)^3}\, e^{ - i {\bf
     q}\cdot {\bf  r}}  \,  G_M (-
     {\bf q}^2 ) \nn [3pt]
   & \equiv &   {\bm
     \nabla_{\bf r}} \times  {\bf M}_{\rm naive} (r) \,,
       \label{rhoint3RN}
\end{eqnarray}
where we employed the condition
imposed on $\tilde\phi$ in Eq.~(\ref{normN}) and used the fact
that the integral over ${\bf \tilde P}$ of an odd function vanishes. 
Here,  $G_E (q^2)$ and $G_M (q^2)$ refer to the Sachs electric and
magnetic form factors \cite{Ernst:1960zza} (note again the unconventional normalization of the Pauli form factor):
\begin{equation}
G_E (q^2) = F_1 (q^2 ) + \frac{q^2}{4 m } F_2 (q^2), \quad \quad \quad
G_M (q^2) = F_1 (q^2 ) + m F_2 (q^2)\,.
\end{equation}
Owing to the classical expression for the magnetization current ${\bf
  J}_{\rm  mag} ({\bf r}) = {\bm \nabla} \times {\bf M} ({\bf r})$, the
vector-valued function  $ {\bf M}_{\rm naive} (r) $ (an operator in
the spin space) is interpreted as the magnetization distribution of
the spin-$1/2$ particle. In the ZAMF of a spherically
symmetric system, the spin operator is the only available
vector, and the function $ {\bf M}_{\rm naive} (r) $ is thus
interpreted as the density of magnetic dipoles. It is
usually expressed in terms of a scalar magnetization density
$\rho_{\rm naive}^{\rm mag} (r)$, which
corresponds to the Fourier transform of the magnetic form
factor $G_M$ in the Breit frame
\begin{equation}
 {\bf M}_{\rm naive} (r) = 
 \frac{1}{2m}  {\bm \sigma} \rho_{\rm naive}^{\rm mag} (r)\,, \quad
 \quad \quad
 \rho_{\rm naive}^{\rm mag} (r) =   \int \frac{d^3 {q}}{(2\pi)^3}\, e^{ - i {\bf
     q}\cdot {\bf  r}}  \, G_M (- {\bf q}^2 )\,.
 \label{RhoMagNaive}
\end{equation}
In the notation we employ, the charge and scalar magnetization densities $\rho_{\rm naive}^{\rm
  ch} (r)$ and  $\rho_{\rm naive}^{\rm mag} (r)$ are normalized to $1$
and  to the magnetic moment $\mu = 1+\kappa$ of the particle,
respectively. These two scalar densities incorporate the complete information about the
internal structure of a composite spin-$1/2$ particle encoded in the
electromagnetic form factors. 

Notice that the  final expressions in Eq.~(\ref{rhoint3RN}) do not depend on the
shape of the wave packet. They correspond to the leading approximation to the
spatial densities for packets localized with $R$ much 
bigger than the Compton wavelength $1/m$ 
while much smaller than any other  length scale characterizing the system. 
Such a definition, however, becomes doubtful for systems like light hadrons, whose characteristic
length scales are comparable to or smaller than the Compton
wavelength, see Ref.~\cite{Jaffe:2020ebz} and references therein. 

\medskip
An alternative definition of spatial densities, applicable to any
systems, is obtained by localizing the wave packet describing the
particle under consideration without performing a nonrelativistic
expansion for the integrand in Eq.~(\ref{rhoint2N}).  
Using the method of dimensional counting, the $R\to 0$
limit in Eq.~(\ref{rhoint2N}) can, in fact,  be taken for arbitrary
(nonvanishing) values of the particle's mass as discussed in
Ref.~\cite{Epelbaum:2022fjc}.
Doing so we obtain
\begin{align}
J^0_{\phi}({\bf r}) & =  \int \frac{d^3 {\tilde P} \, d^3 {q}}{(2\pi)^3} 
\left\{
 F_1\bigg[  \Big( {\bf {\hat{ \tilde P}}\cdot{\bf q} }\Big)^2 - {\bf q}^2\bigg]  
+ \frac{i}{2}\, {\bf q}\cdot \left({\bf\hat{\tilde P} }\times{\bm \sigma} \right)   
 F_2 \bigg[  \Big( {\bf  {\hat{ \tilde P}}\cdot{\bf q} }\Big)^2 - {\bf q}^2\bigg]   \right\}  \, |\tilde\phi({
  {\bf \tilde P}})|^2  \, e^{ - i {\bf q}\cdot {\bf  r}} \,, \nn
{\bf J}_{\phi}({\bf r}) & = \int \frac{d^3 {\tilde P} \, d^3 {q}}{(2\pi)^3}  \left\{ {\bf \hat {\tilde P}} \, F_1\bigg[  \Big( {\bf{\hat{\tilde  P}}\cdot{\bf q} }\Big)^2 - {\bf q}^2\bigg]  
 - \frac{i }{2} \,  \left(  {\bf q}\times{\bm \sigma} - {\bf q}\times{\bf \hat {\tilde P}} \, {\bm \sigma}\cdot{\bf \hat {\tilde P}} \right) 
F_2 \bigg[  \Big({\bf{\hat{\tilde  P}}\cdot{\bf q} }\Big)^2 - {\bf q}^2\bigg]  
 \right\}  
  |\tilde\phi(
  {\bf {\tilde P}})|^2  e^{ - i {\bf q}\cdot {\bf  r}} \,,
\label{rhoint3N}
\end{align}
where ${\bf\hat{\tilde P}}= {\bf{\tilde P}}/\tilde {P}$, with $\tilde {P} \equiv | {\bf \tilde  P}|$.
For spherically symmetric wave packets with $\tilde\phi( {\bf {\tilde
  P}}) =  \tilde\phi(| {\bf {\tilde
  P}}|) $, the integration over ${\bf \tilde P}$  can be easily
performed, and the current densities become independent of $\tilde\phi$. Denoting $\alpha=\cos\theta$ with $\theta$ being the angle  
between the vectors ${\bf{\tilde P}}$ and $\bf{q}$,  we obtain
\begin{eqnarray}
J^0({\bf r}) &  =&  \int \frac{d^3 {q}}{(2\pi)^3}\, e^{ - i {\bf q}\cdot {\bf  r}}
\int_{-1}^{+1} d\alpha\,  \frac{1}{2}\,   F_1\Big[ (\alpha^2-1)
               \,{\bf q}^2\Big] \; \equiv \; \rho_1 ( r)
     \,    , \nn[3pt]
{\bf J} ({\bf r}) & =& \int \frac{d^3 {q}}{(2\pi)^3}\, e^{ - i {\bf q}\cdot {\bf  r}}
\int_{-1}^{+1} \frac{d\alpha}{2}
\, \left\{ \alpha \, {\bf \hat  q}  F_1\Big[ (\alpha^2-1) {\bf q}^2\Big]  
- \frac{i }{4} \,  {\bf q}\times{\bm \sigma}  \left( 1+ \alpha^2 \right) 
F_2\Big[ (\alpha^2-1) \,{\bf q}^2\Big] 
                    \right\} \nn[3pt]
  & = & 
\frac{ {\bm
     \nabla_{\bf r}} \times  {\bm \sigma} }{2m} 
        \int \frac{d^3 {q}}{(2\pi)^3}\, e^{ - i {\bf q}\cdot {\bf  r}}
\int_{-1}^{+1} d\alpha\, \frac{1}{4}
\left( 1+ \alpha^2 \right) 
        m F_2\Big[ (\alpha^2-1) \,{\bf q}^2\Big]\nn[3pt]
  & \equiv &  {\bm
     \nabla_{\bf r}} \times  {\bf M} (r) \; \equiv \;              
       \frac{ {\bm
     \nabla_{\bf r}} \times   {\bm \sigma} }{2m}   \, \rho_2 (r) \,.
\label{rhoint4N}
\end{eqnarray}
Notice that the first term in the curly brackets, being an odd
function in $\alpha$, does not contribute
to the density after integrating over $\alpha$. 
The spatial densities $J^\mu({\bf r})$ defined above
do not depend on the shape of the wave packet and are unambiguously 
expressed in terms of the experimentally measurable form factors $F_1
(q^2)$ and $F_2 (q^2)$. However, in contrast to the naive definition
in Eq.~(\ref{rhoint3RN}), the validity of Eq.~(\ref{rhoint4N}) does not depend on the relationship between the
Compton wavelength $1/m$ and other length scales characterizing the system. 
In particular, Eq.~(\ref{rhoint4N}) can be also used to define the spatial
densities of light handrons.

\medskip 
It is striking that the obtained result for 
$J^\mu ({\bf r})$, expressed in terms of the form factors $F_1(q^2)$ and
$F_2 (q^2)$, does not depend on the particle's mass, similarly to the
charge density for a spinless system introduced in Ref.~\cite{Epelbaum:2022fjc}.  This implies
that the traditional expression for the density, $J^\mu_{\rm naive}
({\bf r})$, does \emph{not}
emerge from $J^\mu ({\bf r})$ by expanding about the static limit.
As explained in Ref.~\cite{Epelbaum:2022fjc} for the case of a spin-$0$ particle, 
this counterintuitive feature originates from 
the non-commutativity of the $R \to 0$ and $m \to \infty$ limits
of $J^\mu_\phi ({\bf r})$ in Eq.~(\ref{rhoint2N}).
Notice that while a finite-order approximation in the $1/m$-expansion is valid
when calculating the form factors in Eq.~(\ref{eqMEN}) provided $-q^2 \ll m^2$, its validity is
violated in certain momentum regions when performing the integration in Eq.~(\ref{rhoint2N})
if $R$ is taken  of the order of the Compton wavelength or smaller.  

\medskip
The ZAMF expression for the charge density $\rho_1 (r)$
in Eq.~(\ref{rhoint4N}) 
coincides with that of a scalar particle discussed in
Ref.~\cite{Epelbaum:2022fjc}. Given that the state $| \Phi, {\bf 0}, s \rangle$ is
an eigenstate of the charge operator with an eigenvalue 
$1$, both  $\rho_1 (r)$ and the naive charge density $\rho_{\rm
  naive}^{\rm ch} (r)$ are normalized to $1$. On the other hand, 
$\rho_2 (r)$ is normalized to $\frac{2}{3} \kappa$ rather than to the
magnetic moment $1 + \kappa$ as it is the case for the naive
magnetization density $\rho^{\rm mag}_{\rm naive} (r)$.
We will come back
to this point  in section \ref{moving}, where a geometrical interpretation of the obtained rest-frame
densities will be given in terms of the densities defined  in the infinite-momentum
frame. To facilitate such an interpretation, 
it is instructive to rewrite $J^\mu ({\bf r})$, specified
in Eq.~\eqref{rhoint4N}, as 
\be
\label{RhoCoordIndepN}
J^\mu ({\bf r}) = \frac{1}{4 \pi} \int d {\bf \hat n} \, J^\mu_{{\bf \hat n}} ({\bf r})\,,
\ee
where ${\bf \hat n} \equiv {\bf n}/ | {\bf n}|$ is a unit vector along
the direction of the vector ${\bf \tilde P}$ in Eq.~(\ref{rhoint3N}) and 
\begin{equation}
  \label{RhoCoordIndepN1}
J^0_{{\bf \hat n}}({\bf r}) = \rho_{1, {\bf \hat n}} (
{\bf r} )\, , \quad \quad \quad
{\bf J}_{{\bf \hat n}} ({\bf r}) =  \frac{1}{2m}  
                                    {\bm \nabla}_{\bf r} \times  {\bm\sigma}_\perp  \, 
     \rho_{2, {\bf \hat n}} ( {\bf r} ) 
\,, 
\end{equation}  
where
\begin{equation}
  \label{AuxDens}
\rho_{i, {\bf \hat n}} ( {\bf r} ) = \rho_{i} ( {r_\perp}
)  \,  \delta ( r_\parallel ) \,  , 
\end{equation}
and the two-dimensional  auxiliary densities $\rho_{i} ( {r_\perp})$ are defined in terms of the form factors
$F_i (q^2)$ via
\begin{equation}
  \label{RhoCoordIndepN2}
\rho_{1} ( {r_\perp} )  =
\int \frac{d^2 {q}_\perp}{(2\pi)^2}\, e^{ - i {\bf q}_\perp \cdot {\bf  r}_\perp}\, 
 F_1\left(-{\bf q}_\perp^2\right) \,,
\quad \quad \quad
\rho_{2} ( {r_\perp} )  =
\int \frac{d^2 {q}_\perp}{(2\pi)^2}\, e^{ - i {\bf q}_\perp\cdot {\bf  r}_\perp}\, 
m  F_2\left(-{\bf q}_\perp^2\right) \,.
\end{equation}
Here and in what follows, ${\bf a}_\parallel \equiv {\bf a} \cdot
{\bf \hat n} \, {\bf \hat n} $
(${\bf a}_\perp \equiv {\bf a} -  {\bf a} \cdot
{\bf \hat n} \, {\bf \hat n} =
{\bf \hat n}
\times ({\bf a} \times {\bf \hat n} )$) denote the component of a
vector ${\bf a}$ parallel (perpendicular) to the vector $\bf n$, and
$a_\parallel \equiv | {\bf a}_\parallel |$,  $a_\perp \equiv | {\bf
  a}_\perp |$.
Notice that the auxiliary densities $\rho_{i, {\bf \hat n}} ( {\bf r}
)$ depend on the direction ${\bf \hat n}$ only through the arguments
$r_\perp$ and $r_\parallel$
of the corresponding two-dimensional densities and the
delta-function.

\section{The electromagnetic densities of a spin-1/2 system in moving frames}
\label{moving}

To generalize the expressions in Eqs.~(\ref{rhoint4N}) and (\ref{RhoCoordIndepN}) to moving frames,
we follow the procedure of Ref.~\cite{Epelbaum:2022fjc} and replace the packet
 in Eq.~(\ref{rhoint2N})
  with its boosted expression. 
Using
Eq.~(\ref{statedefN2}) and the transformation
properties of the momentum eigenstates under a boost $\Lambda_{\bf v}$ with velocity
${\bf v}$,
\begin{equation}
  |p, s \rangle  \; \stackrel{\Lambda_{\bf v}}{\longrightarrow} \; 
U (\Lambda_{\bf v} ) |p, s \rangle = \sum_{s_1} D_{s_1 s} \bigg[
W\bigg( \Lambda_{\bf v}, \frac{{\bf p}}{m} \bigg) \bigg]
|\Lambda_{\bf v} p, s_1 \rangle\,,
\end{equation}
where $D_{s_1s}\left[ W\right]$ is a spin-1/2 representation of the
corresponding Wigner rotation $W$  \cite{Weinberg:1995mt}, 
we express 
 a normalizable Heisenberg-picture state  located at the origin of a moving frame in terms of the spherically symmetric ZAMF
quantity $\phi ({\bf p})$ as (see also Ref.~\cite{Hoffmann:2018edo}):   
\begin{equation}
|\Phi, {\bf 0},s \rangle _{\bf v}  =  \int \frac{d^3 {p}}{\sqrt{2 E (2\pi)^3}}  \, \sqrt{\gamma \Big( 1-\frac{{\bf v} \cdot {\bf
                                        p}}{E} \Big)} \, \phi \Big(
                                        \Lambda_{\bf v}^{-1} {\bf p} \Big)
  \sum_{s_1}D_{s_1s}\bigg[ W\bigg( \Lambda_{\bf v}, 
    \frac{ \Lambda_{\bf v}^{-1} {\bf p} }{m} \bigg) \bigg]  |p ,s_1 \rangle,  
\label{statedefN2Moving}
\end{equation} 
where $\gamma = (1 - v^2)^{-1/2}$,
$E = \sqrt{m^2 + {\bf p}^2}$, $\Lambda_{\bf v}$ is the Lorentz boost
from the ZAMF to the moving frame and $\Lambda_{\bf v}^{-1} {\bf p} =   {\bf {\hat v}} \times ({\bf p}
\times   {\bf {\hat v}}) + \gamma ({\bf p} \cdot  {\bf {\hat v}} - v E )  {\bf {\hat  v}}$.

Analogously to the case of the ZAMF, we use the method of dimensional
counting by substituting ${\bf P} = {\bf \tilde  P}/R$ and obtain in the $R \to 0$ limit: 
\begin{align}
J^0_{\phi,  {\bf v}}({\bf r}) & =  \int \frac{d^3 {\tilde P} \, d^3 {q}}{(2\pi)^3} 
\, \gamma \left(1  - {\bf v} \cdot {\bf\hat{\tilde  P}} \right)  
 \left\{
 F_1\bigg[  \Big( { {\bf \hat{\tilde P}}\cdot{\bf q} }\Big)^2 - {\bf q}^2 \bigg]   \right. \nn  & \left. 
+  \frac{1}{4}\left( 
  {\bf q} \cdot {\bm \Sigma}_{{\bf v,  {\bf \hat m}}}  \, {\bf \hat{\tilde P} }\cdot {\bm\sigma} -
                                                                                                  {\bf
                                                                                                  \hat{\tilde
                                                                                                  P}
                                                                                                  }\cdot
                                                                                                  {\bm\sigma}
                                                                                                  \,
                                                                                                  {\bf
                                                                                                  q} \cdot {\bm \Sigma}_{{\bf v, \hat {\bf m}}}  
 \right)
F_2\bigg[  \Big( { {\bf \hat{\tilde P}}\cdot{\bf q} }\Big)^2 - {\bf q}^2 \bigg]    \right\} \, \big|\tilde\phi \big({\bf \tilde P}' \big)
\big|^2  \, e^{ - i {\bf q}\cdot {\bf  r}}\,, \nn
{\bf J}_{\phi,  {\bf v}}({\bf r}) & = \int \frac{d^3 {\tilde P} \, d^3 {q}}{(2\pi)^3}\, 
\gamma \left(1  - {\bf v} \cdot {\bf  \hat{\tilde P}} \right)  
  \left\{ 
  \frac{1}{2}\left( 
{\bm  \Sigma }_{{\bf v, \hat {\bf m}}}
\, {\bf \hat{\tilde P} }\cdot {\bm\sigma} +
                                    {\bf \hat{\tilde P} }\cdot
                                    {\bm\sigma} \,  {\bm
                                    \Sigma}_{{\bf v, \hat {\bf m}}}
                                    \right)  F_1\bigg[  \Big( {\bf { \hat{\tilde P}}\cdot{\bf q} }\Big)^2 - {\bf q}^2 \bigg]    \right. \nonumber\\ & \left.  
- \frac{i  }{4} \, {\bm  q}  \times  
\left( 
  {\bm  \Sigma }_{{\bf v, {\bf \hat m}}}  -  {\bf \hat{\tilde P} }\cdot {\bm\sigma}
                                                                                                                                             \,   {\bm  \Sigma}_{{\bf v, \hat {\bf m}}}  
\, {\bf \hat{\tilde P} }\cdot {\bm\sigma}  \right) 
F_2\bigg[  \Big( {\bf  { \hat{\tilde P}}\cdot{\bf q} }\Big)^2 - {\bf q}^2 \bigg]  
 \right\} 
   \,
\big|\tilde\phi \big({\bf \tilde P}' \big)
\big|^2 
 \, e^{ - i {\bf q}\cdot {\bf  r}}\,,
\label{rhoint3boostedN}
\end{align}
where we have introduced ${\bf \tilde P}' =
 {{\bf \hat  v}} \times \big( {\bf \tilde P} \times  {{\bf \hat  v}}\big)   +
  \gamma \big({\bf  \tilde  P}  \cdot  { {\bf \hat   v}}   -   v \tilde
  P\big)    {{\bf \hat  v}}$ and a unit vector ${{\bf \hat  m}} \equiv 
  {\bf  {\hat{\tilde P}}'}$. Furthermore,  ${\bm  \Sigma }_{{\bf v, \hat
      {\bf m}}}$  refers to the Wigner rotated spin-operator
\begin{equation}
 {\bm  \Sigma }_{{\bf v, 
      {\bf \hat m}}} \equiv D^\dagger \big[ W \big( \Lambda_{\bf v}, 
{{\bf  \hat m}} \big) \big]  {\bm  \sigma } D \big[ W \big( \Lambda_{\bf v}, 
{{\bf  \hat m}} \big) \big] \,,
\end{equation}
with 
\begin{equation}
D \big[ W \big( \Lambda_{\bf v}, 
{{\bf  \hat m}} \big) \big] = \lim_{R \to 0} D \bigg[ W \bigg(
 \Lambda_{\bf v}, \frac{\Lambda_{\bf v}^{-1} ({\bf \tilde P}/R)}{m}
 \bigg) \bigg] \,.
\end{equation}
Next, we change the integration variable ${\bf  \tilde P} \to {\bf \tilde P}'$ and define a vector valued function
\begin{equation}
  {\bf n} \big({\bf v},  { {\bf \hat m}} \big) = { {\bf \hat v}} \times \big( {\bf  \hat m} \times  { {\bf \hat v}}\big)
+ \gamma \big( {\bf  \hat m} \cdot  { {\bf \hat v}}  + v )  {{\bf \hat v}} \,.
\end{equation}
Given that ${\bf \tilde P} =
 {{\bf \hat  v}} \times \big( {\bf \tilde  P}' \times  {{\bf \hat v}}\big)   +
  \gamma \big( {\bf \tilde  P}'  \cdot  { {\bf  \hat  v}}   +   v \tilde
  P'\big)    {{\bf \hat v}}$, it follows that  $ { {\bf \hat n}} = {\bf  {\hat {\tilde P}}} $.  
It is easy to  verify that the Jacobian of the change of variables ${\bf \tilde P} \to {\bf\tilde P}'$
cancels the first factor in the integrands in Eq.~(\ref{rhoint3boostedN}), leading to
\begin{eqnarray}
  \label{rhoint3boosted2}
J^0_{\phi,  {\bf v}}({\bf r}) & =  & \int  \frac{d {
                                     {\bf \hat m}} \, d\tilde {P}' \, d^3
  {q}}{(2\pi)^3}
                                   \; \big|\tilde\phi \big(
                                     {\bf \tilde P} ' \big) \big|^2  \, e^{ - i {\bf q}\cdot {\bf  r}} \bigg\{
 F_1\Big[ ({\bf \hat n}\cdot {\bf q}) ^2  - {\bf
                q}^2 \Big]  
             +
\frac{1}{4} \left( 
 {\bf {q} }\cdot {\bm\Sigma}_{{\bf v, \hat {\bf m}}} \, 
{\bf \hat{n} }\cdot {\bm\sigma} - {\bf \hat{n} }\cdot {\bm\sigma}  \,   {\bf {q} }\cdot {\bm\Sigma}_{{\bf v, \hat {\bf m}}} 
\right)
F_2\Big[ ({\bf \hat n}\cdot {\bf q}) ^2  - {\bf
                q}^2 \Big]   \bigg\}  
,  \nonumber \\
{\bf J}_{\phi,  {\bf v}}({\bf r}) & = & \int \frac{d {
                                     {\bf \hat m}} \,  d \tilde {P}' \, d^3
  {q}}{(2\pi)^3}
                                   \; \big|\tilde\phi \big(
\tilde {\bf P} ' \big) \big|^2  \, e^{ - i {\bf q}\cdot {\bf  r}}  
\bigg\{ 
\frac{1}{2} \left( {\bm  \Sigma}_{{\bf v, {\bf \hat m}}}
\, {\bf \hat{n} }\cdot {\bm\sigma} + {\bf \hat{n} }\cdot {\bm\sigma} \, {\bm  \Sigma}_{{\bf v, \hat {\bf m}}}  
\right)  F_1\Big[  \left({{\bf \hat n}}\cdot{\bf q} \right)^2 - {\bf q}^2\Big]  \nonumber\\ & - &  
 \frac{i  }{4} \, {\bm  q}  \times  
\left( 
 {\bm  \Sigma}_{{\bf v, \hat {\bf m}}}  -  {\bf \hat{n} }\cdot
                                                                                                 {\bm\sigma} \, {\bm  \Sigma}_{{\bf v, \hat {\bf m}}}  
\, {\bf \hat{n} }\cdot {\bm\sigma}  \right) 
F_2\Big[  \left({ {\bf \hat n}}\cdot{\bf q} \right)^2 - {\bf q}^2 \Big] 
\bigg\} .
                        \nonumber
\end{eqnarray}
Using the spherical symmetry of $\tilde\phi \big(
{\bf \tilde P} ' \big) = \tilde\phi \big(|
{\bf \tilde P} '| \big) $ and Eq.~(\ref{normN}), the integration over
$\tilde P'$ becomes trivial, yielding the densities  which are independent
of the wave packet form. To keep the notation compact, we express the
resulting densities in the form similar to Eq.~(\ref{RhoCoordIndepN}):
\be
\label{RhoCoordIndepNv}
J^\mu_{\bf v} ({\bf r}) = \frac{1}{4 \pi} \int d {\bf \hat m} \,
 J^\mu_{{\bf v},  {\bf \hat m}} ({\bf r})\,,
\ee
where
\begin{eqnarray}
  \label{RhoCoordIndepNv2}
J^0_{{\bf v}, {\bf \hat m}} ({\bf r}) &=&   \rho_{1,  {\bf \hat n}} ( {\bf r})  +    
                                         \frac{i}{4 m}   {\bm
                                   \nabla}_{\bf r} \cdot \left[
                                          {\bm\Sigma}_{{\bf v, 
                                          {\bf \hat m}}} , 
\; {\bf \hat{n} }\cdot {\bm\sigma} \right]_-
                                         \rho_{2, {\bf \hat n}} ( {\bf
                                         r} ) , 
     \nn [3pt]
{\bf  J}_{{\bf v}, {\bf \hat  m}} ({\bf r}) &=& \frac{1}{2}
 \left[ 
 {\bm  \Sigma }_{{\bf v, {\bf \hat m}}} , \; {\bf \hat{n} }\cdot {\bm\sigma}  \right]_+  \rho_{1, {\bf  \hat n}} ( {\bf r}) + 
 \frac{1 }{4 m} \, {\bm \nabla}_{\bf r}   \times  \left(  \left[
                                          {\bm\Sigma}_{{\bf v,
                                          {\bf  \hat m}}} , 
\; {\bf \hat{n} }\cdot {\bm\sigma} \right]_-
\,  {\bf \hat{n} }\cdot {\bm\sigma} \right) 
 \rho_{2,  {\bf \hat n}} ( {\bf r}) 
     \,,     
\end{eqnarray}
with $[ A, \; B ]_\pm \equiv AB \pm BA$ and the auxiliary densities  $\rho_{i, {\bf \hat n}} ( {\bf r} )$
being defined in Eq.~(\ref{AuxDens}) with 
the superscripts
$\perp$ and $\parallel$ of various vectors referring, as before, to 
the direction of ${\bf n}$.   

The above expressions simplify considerably in two extreme
cases. First, in the particle's ZAMF with $v = 0$ and $\gamma =
1$, we have $  {\bf n} \big({\bf v},  {{\bf \hat m}} \big) = {\bf \hat 
  m}$  and $ {\bm  \Sigma}_{{\bf v, {\bf \hat m}}} = {\bm \sigma} $. Thus, one can
simply replace in Eq.~(\ref{RhoCoordIndepNv}) ${\bf \hat m}$ with
${\bf \hat n}$. 
Then,  given Eq.~(\ref{AuxDens}),
the integration over ${\bf \hat n}$ of the second (first) term on the
right-hand side of the first (second) equality in
Eq.~(\ref{RhoCoordIndepNv2}) yields a vanishing result, so that one
encounters the expressions in Eqs.~(\ref{RhoCoordIndepN})-(\ref{RhoCoordIndepN2}). 

The second interesting case corresponds to the infinite-momentum frame
(IMF) with $v \to 1$ and $\gamma
\to \infty$, for which the vector-valued function ${\bf \hat  n} $ turns to $
{\bf \hat v}$. Then,  the integrand in Eq.~(\ref{RhoCoordIndepNv})  depends on 
${\bf \hat m}$ only through the Wigner rotation matrices. Performing the integration over 
${\bf \hat  m}$, we obtain for the spatial densities in the IMF
\begin{eqnarray}
  \label{RhoIMF1}
  J^0_{\rm IMF} ({\bf r}) &\equiv& J^0_{{\bf \hat v}} ({\bf r}) \; =
                                    \; 
                                           \rho_{1,  {\bf \hat v}} ( {\bf
                                         r}) \;  + \;   {\bf \hat v} \cdot \,
                                         \frac{   {\bm \nabla}_{\bf r} \times {\bm\sigma}_\perp  }{4m}   \, 
                                         \rho_{2, {\bf \hat v}} ( {\bf
                                         r} ) \,, \nn[3pt]
                                   {\bf J}_{\rm IMF} ({\bf r}) &\equiv &
  {\bf J}_{{\bf \hat  v}}  ({\bf r}) 
  \; =\; 
                                         \frac{   {\bm \nabla}_{\bf r} \times {\bm\sigma}_\perp  }{4 m}   \, 
                                         \rho_{2, {\bf \hat v}} ( {\bf r} ) \,,     
\end{eqnarray}
where the densities $\rho_{i, {\bf \hat v}} ( {\bf r} ) $ are given in
Eqs.~(\ref{AuxDens}) and (\ref{RhoCoordIndepN2}) with ${\bf \hat  n} =
{\bf \hat v}$. 
The appearance of  $\delta
(r_\parallel )$, see Eq.~(\ref{AuxDens}), reflects the fact that in
the IMF, the system is Lorentz-contracted to a two-dimensional object
perpendicular to the velocity ${\bf \hat  v}$ of the moving frame. It
thus makes sense to introduce two-dimensional densities in the
impact parameter  space spanned by ${\bf r}_\perp$ by integrating the
three-dimensional densities in Eq.~(\ref{RhoIMF1}) over $r_\parallel$:
\begin{eqnarray}
  \label{RhoIMF12dim}
\int d r_\parallel \,  J^0_{{\bf \hat v}} ({\bf r})  &=& \rho_1 (r_\perp
                                                      )  \; + \;   {\bf \hat v} \cdot \,
                                         \frac{ {\bm \nabla}_{\bf r}  \times   {\bm\sigma}_\perp }{4m}   \, 
                                         \rho_{2}  ( r_\perp) \,, \nn[3pt]                           
\int d r_\parallel \,  {\bf J}_{{\bf \hat  v}}  ({\bf r}) &=& 
                                                      \frac{    {\bm \nabla}_{\bf r} \times {\bm\sigma}_\perp
                                                      }{4 m}   \, 
                                         \rho_{2}  ( r_\perp) \,. 
\end{eqnarray}
Notice that the two-dimensional transverse charge and magnetization densities 
of spin-1/2 particles in the IMF have been extensively discussed in the
literature, see e.g.~\cite{Burkardt:2002hr,Miller:2007kt,Carlson:2007xd}. 
As for the expression in the last line of 
Eq.~(\ref{RhoIMF12dim}), the quantity $\frac{1}{2m} \rho_{2}  (
r_\perp) $ 
has been interpreted in Ref.~\cite{Miller:2007kt} as a true magnetization density
of the system, which generates the anomalous magnetic moment. On the other hand, Ref.~\cite{Miller:2010nz} argued that a
  more natural interpretation of the anomalous magnetization
  density is provided by the distribution
  \be
\tilde \rho_2 ({\bf r}_\perp) = -\frac{1}{2m} r_y \frac{\partial \rho_{2}  (
r_\perp)}{\partial r_y}  \,,
  \ee
where $r_y$ is a component of ${\bf r}$ that is perpendicular to both ${\bf
  v}$ and $ {\bm\sigma}_\perp$. Clearly, both densities $\frac{1}{2m}  \rho_{2}  (
r_\perp) $ and $\tilde \rho_2 ({\bf r}_\perp)$ are
normalized to the anomalous magnetic moment of the particle.

\section{Discussion and interpretation}
\label{sec:Disc}

Before discussing the interpretation of the obtained results, it
is useful to briefly summarize what has been achieved so far. Starting from a
general matrix element of the electromagnetic current operator $\hat
j^\mu  ({\bf r}, 0 )$ in a wave-packet state $ | \Phi, {\bf 0},s \rangle $ of  a
composite spin-$1/2$ system placed at the origin of the ZAMF, 
we have taken the limit of a sharp localization, $R
\to 0$, in order to remove the information about the shape of the wave packet 
(for a spherically symmetric $\phi (| {\bf p} |)$), thereby obtaining the
corresponding spatial density $J^\mu ({\bf r})$. The resulting
expression in the first line of Eq.~(\ref{rhoint4N}) relates the charge
density
$J^0 (r) \equiv \rho_1 (r)$ to the Dirac form factor $F_1 (q^2)$, while
the current distribution ${\bf J} ({\bf r})$ is expressed in terms of
the scalar magnetization density $\rho_2 (r)$ related to
the Pauli form factor $F_2 (q^2)$. These expressions for the 
charge and current densities differ from the conventional ones given
in terms of the Fourier transform of the Sachs form factors in the Breit frame,
see Eqs.~(\ref{rhoint3RN}) and (\ref{RhoMagNaive}).  This raises
questions about the uniqueness of our results, their relation to the
conventional densities and  
physical interpretation,
which will be addressed below.
To keep the
discussion as transparent as possible, we focus in the following
on a spin-$0$ system considered in Ref.~\cite{Epelbaum:2022fjc} (but the arguments
hold for spin-$1/2$ particles as well). The 
charge densities $\rho (r)$  and $\rho_{\rm naive} (r)$ of Ref.~\cite{Epelbaum:2022fjc} 
for a spinless system
coincide with $\rho_1 (r)$  and $\rho_{\rm naive}^{\rm ch} (r)$ of
this paper if one sets $F_1 (q^2) \equiv F (q^2)$ and $F_2 (q^2) = 0$. 

To uncover the meaning of the charge distribution we
need to identify the way it can be probed experimentally, and we
usually think of elastic lepton scattering.
For an infinitely heavy system described by a 
 {\it static} charge distribution $\rho_{\rm naive} (r)$,
 there are no recoil effects, and the differential cross section
 in the single-photon approximation is given
by the  Mott cross section for scattering off a point-like charge times
$| F ( - {\bf q}^2 ) |$, see e.g.~\cite{Halzen:1984mc} for a textbook discussion.  
This exact result provides the means for directly accessing $\rho_{\rm naive} (r)$ 
experimentally, and it runs deep in our way of visualizing
and interpreting the charge distribution of a composite system
in nonrelativistic, quantum mechanical  settings as common in atomic, nuclear
and solid-states physics. Notice that the static limit actually
corresponds to the limit of $c \to \infty$, in which the Lorentz group reduces
to the Galilean one, so that the charge distribution $\rho_{\rm naive}
(r)$ becomes frame independent. 

Clearly, the static limit is merely an idealization that can be
imposed to approximate low-energy dynamics of composite systems. 
Contrary to what is sometimes claimed in the literature, see
e.g.~\cite{Halzen:1984mc}, the Breit distribution $\rho_{\rm naive}
(r)$ can {\it not} be interpreted as the intrinsic charge density of
the system beyond the strict static limit \cite{Epelbaum:2022fjc}. While it is possible
to perturbatively (i.e., based on the $1/m$-expansion) take into account
corrections beyond the static limit using an alternative definition of
the  charge density with the wave packet localized 
at distances well above its Compton
wavelength \cite{Jaffe:2020ebz}, the resulting spatial distribution does not
entirely reflect the internal structure of the
system, being dependent on the wave packet. 
The usefulness of such a definition of the charge
density thus relies on the Compton wave length being much
smaller than the size of the system as determined by the charge radius
\cite{Jaffe:2020ebz}. 

In contrast, the charge density defined using sharply
localized wave packets as done here and in Ref.~\cite{Epelbaum:2022fjc} is completely determined
by intrinsic properties of the system, regardless of
the relationship between the Compton wave length and the
characteristic scale of the system related to the charge
radius.  Moreover, the resulting charge distribution $\rho (r)$ comes
out to be independent of the particle's mass, which implies that $\rho_{\rm naive}
(r) \neq \lim_{m \to \infty}\rho (r)$ and makes the static picture
described above inappropriate for the interpretation of $\rho
(r)$. More precisely, the large momentum components of the wave packet
play a crucial role in the definition of the charge density $\rho
(r)$, which thus represents an intrinsically relativistic quantity.  
Remarkably, imposing the sharp localization limit in the definition of
the charge density forces one to think of $\rho (r)$  in a
holographic-like picture in terms of
a continuous superposition of images taken in all possible
IMF \cite{Epelbaum:2022fjc}. The usual non-relativistic interpretation
of the rest-frame charge distribution of a heavy system characterized by the
density $\rho (r)$ can only be obtained by reconstructing $\rho_{\rm
  naive} (r)$ from $\rho (r)$. 

The above-mentioned holographic-like relationship between the ZAMF and
IMF distributions also holds for the electromagnetic
densities of a spin-1/2 particle considered in this paper. 
Comparing Eqs.~(\ref{RhoCoordIndepN}) and (\ref{RhoCoordIndepN1}) to
Eq.~(\ref{RhoIMF1}), one observes that the spatial current densities
$J^\mu ({\bf r})$ in the ZAMF are given by integrating the
three-dimensional IMF densities $J^\mu_{\hat {\bf v}} ({\bf r})$ over all possible
directions:
\begin{equation}
J^0 ({\bf r})  =  \frac{1}{4 \pi} \int d {\bf \hat  v} \, J^0_{{\bf \hat
    v}} ({\bf r})\,,\quad \quad
{\bf J} ({\bf r})  =  2\times \frac{1}{4 \pi} \int d {\bf \hat  v} \, {\bf J}_{{\bf \hat v}} ({\bf r})\,.
\label{intvIMF}
\end{equation}
This feature can be understood already by looking at the defining expressions for the densities in
Eq.~(\ref{rhoint2N}): taking the limit $R \to 0$ to remove the
information about the wave packet profile function from the definition
of the densities brings the integrands in Eq.~(\ref{rhoint2N}) to the
kinematics with the total momentum $| {\bf P} |$ being larger than all other
momentum scales, which in turn corresponds to the IMF (up to the
Wigner rotations). In the IMF, the system is Lorentz-contracted to essentially a two-dimensional
object perpendicular to the velocity of the moving frame with the
densities given in Eq.~(\ref{RhoIMF1}). While only such
two-dimensional images of the system are observed in the IMF, the expression
in the ZAMF reconstructs the full three-dimensional structure by
putting together all possible two-dimensional ``images''. In general, the
full image of a $d$-dimensional
object can be reconstructed by putting together its 
all possible $d-1$ dimensional ``images''. This kind of representation is possible for all positive integers $d>1$.      
We further emphasize that the extra factor of $2$ that appears in the
second equality in Eq.~(\ref{intvIMF}) has its origin in the Wigner rotations 
of spin states when performing Lorentz boosts to the IMF. 

It is worth mentioning that
the second moment of the
charge distribution $\rho (r)$, the quantity that should be interpreted as the
mean square charge radius of the system, is related to the form factor
slope via
\begin{equation}
\label{eq:newrad}
\langle r^2 \rangle = 4 F_1' (0)~,
\end{equation}
where the prime denotes differentiation with respect to $q^2$. This is in contrast to the usual
relationship
\begin{equation}
\langle r^2 \rangle_{\rm naive} = 6 \left(F_1' (0) +\frac{F_2(0)}{4 m}\right)~, 
\end{equation}
motivated by the static definition $\rho^{\rm ch}_{\rm naive} (r)$. Thus, the  size of e.g.~the proton measured
by the new charge distribution $\rho_{1} (r)$  is $\sqrt{\langle r_{\rm p}^2 \rangle} =
0.62649$~fm, which differs from  $\sqrt{\langle r_{\rm p}^2 \rangle_{\rm naive}} =
0.8409(4)$~fm \cite{ParticleDataGroup:2020ssz,Lin:2021xrc}.
This discrepancy, however, has no practical implications since the 
radius extracted from electron-proton scattering as well as from electronic
and muonic hydrogen is {\it defined} based on the expansion of the
electric Sachs form factor
around $q^2 =0$,
\begin{equation}
G_E (q^2)=G_E (0)\left[1 +q^2 \frac{\langle r^2 \rangle}{6} +\ldots \right]\, ,
\end{equation}  
which is consistently used in the theory underlying these processes.
What we have shown is that the radius directly related to the internal charge density, defined via sharply localized packets,
is indeed smaller due to the squeezing of the density already explained in
Ref.~\cite{Epelbaum:2022fjc}. This also applies to the magnetic
radius related to the slope of the Pauli form factor at zero momentum transfer.

\section{Summary and conclusions}
\label{summary}

Using the prescription introduced in Ref.~\cite{Epelbaum:2022fjc}, we worked out the details
of an unambiguous definition of spatial distributions
of the electromagnetic current for composite spin-$1/2$ systems. 
We obtained relationships between the form factors and the local densities in the ZAMF and moving
frames. The resulting spatial densities are independent of the
specific form of the wave packet in which the state was prepared, provided that it is
spherically symmetric and independent of the spin polarization of the state in the ZAMF of the system. 
Our definition is applicable to any system, irrespective of the relation
between the Compton wavelength and other length scales characterizing
the system. We have also worked out the expressions for the
electromagnetic densities in moving frames including IMF.
Similarly to the charge density of a spin-0 system studied in
Ref.~\cite{Epelbaum:2022fjc}, the obtained electromagnetic spatial distributions
possess a holographic-like interpretation in terms of the
two-dimensional ``images'' made
in all possible IMF. 

We also explored an alternative way to define the spatial densities by
employing  the static approximation as suggested in
Ref.~\cite{Jaffe:2020ebz}, thereby recovering the conventional
expressions in terms of the Fourier transform of the Sachs form
factors in the Breit frame. 
Comparing the two definitions, we find that the static distributions 
can {\it not} be obtained as a systematic approximation to our exact expressions. This is due
to non-commutativity of the limits of an infinitely heavy system and a
sharply localized packet.

\acknowledgements
We thank C\'edric Lorc\'e for useful comments on the manuscript. 
This work was supported in part by BMBF (Grant No. 05P18PCFP1), by
DFG and NSFC through funds provided to the Sino-German CRC 110
“Symmetries and the Emergence of Structure in QCD” (NSFC Grant
No. 11621131001, DFG Project-ID 196253076 - TRR 110),
by ERC  NuclearTheory (grant No. 885150) and ERC EXOTIC (grant No. 101018170),
by CAS through a President’s International Fellowship Initiative (PIFI)
(Grant No. 2018DM0034), by the VolkswagenStiftung
(Grant No. 93562), and by the EU Horizon 2020 research and
innovation programme (STRONG-2020, grant agreement No. 824093).

\end{document}